# Spin and charge distributions in Graphene/Nickel (111) substrate under Rashba spin-orbital coupling

*C.H.Wong\* & A.F.Zatsepin*

*Institute of Physics and Technology, Ural Federal University, Russia*

ch.kh.vong@urfu.ru

Abstract:

To understand the coupling factor between Rashba spin-orbital interaction and ferromagnetic proximity effect, we design a Monte Carlo algorithm to simulate the spin and charge distributions for the room-temperature Rashba material, Graphene/Nickel(111) substrate, at finite temperature. We observe that the rate of exchange fluctuation is a key player to produce giant Rashba spin-orbit splitting in graphene. More importantly, we monitor the Rashba spin-splitting phenomenon where the spin-polarized electrons may be escaped from two opposite edges upon heating. However, the escaped electrons show Gaussian-like distribution in interior area that is important for spintronic engineers to optimize the efficiency of spin-state detection. In addition, we investigate if our Monte Carlo model can explain why room-temperature Rashba effect is observed in Graphene/Nickel(111) substrate experimentally. All results are presented in physical units.

Introduction:

Spin-orbital coupling is a crucial parameter to control the performance of spintronic devices [1,2]. The modern spintronic devices capable of manipulating the dynamic and the spin state of electron without the use of external magnetic field have drawn enormous attention from community [3]. The early spintronic devices were made up of heavy elements owing to the scientific phenomenon that the strength of intrinsic spin-orbital coupling increases with the atomic number of element [4]. However, worldwide scientists have been investigated if carbon-based nanomaterials can trigger spin-orbital coupling at reasonable strength [5,6]. Despite the atomic number of carbon atom is as low as 6, the two dimensional structure of graphene, may be a promising material for spintronic applications due to their exceptionally long diffusion length of spin (~100μm) [7]. However, the extremely long diffusion length of spin is originated from weak spin-orbit coupling [7] and that's why the structure of graphene has to be modified in order to produce robust performance in spintronic.

Ferromagnetic proximity effect is proven to be an effective ingredient to amplify Rashba effect [8,9,10]. When graphene is deposited on nickel (111) substrate, the Rashba-type spin-orbital coupling of the composite is surprisingly strong [9]. While Rashba spin-orbital interaction gives rise to new scientific phenomena such as unconventional spin Hall effect, it has led to another hot research topic of topological insulators [10]. In the presence of Rashba spin-orbital interaction, there is always a momentum-dependent splitting of spin under an external electric field perpendicular to the 2D plane [3,10]. The spin-splitting phenomenon without the need of external magnetic field in Rashba system is credited to strong spin–orbit coupling and asymmetric potential of crystal [3,10]. However, the spin-hall conductivity becomes very weak upon heating [11]. Developing room-temperature Rashba material has been a challenging problem especially for carbon-based materials [6].

To concern the above issues, we design a new Monte Carlo algorithm which allows researchers to understand the coupling factor between Rashba spin-orbital coupling and ferromagnetic proximity interaction. On the other hand, the size of Rashba-type spintronic devices has been considered in order to meet industrial outcomes [12]. A precise control of the trajectory and the spin state of electron are necessary to fabricate state-of-the-art spintronic devices and therefore we are going to examine how the trajectory of electron changes in finite space when temperature goes up.

Computational algorithm:

The size of square lattice is 64 x 64 where $i$ and $j$ refer to the number of grid. The Hamiltonian is listed below.

$$E = \sum_{i,j} \left[ \frac{\hbar^2 k^2 R^2(i,j)}{2m^*} + E_{Rashba} + J_0 e^{-\frac{|r(i,j)-r(i',j')|}{r_o}} S_{i,j} \cdot S_{i'j'} \right]$$

The wavevector of electron within XY plane is $k = \sqrt{k_x^2 + k_y^2}$. $\hbar$ and $m^*$ are the Planck constant and the effective mass of electron, respectively. $R(i,j)$ is the random number between 0 and 1 so that the effect of quantum fluctuation is included in k-space. $J_0$ is the exchange energy [13]. The exchange interaction is covered up to four nearest neighbors. The distance between electrons is marked by $r(i,j) - r(i',j')$. The average separation between electron $r_a$ is converted from the Fermi energy. The spin-up $S_+$ and spin-down $S_+$ state of electron are defined as $S_\pm = \pm \frac{\hbar}{2}$ [14]. In our model, the spin-polarized electrons with the effective mass $m^*$ are suffered from random collision and hence the Rashba potential is approximated as

$$E_{Rashba}(i,j) = \alpha k R(i,j)\delta(S)$$

where $\alpha$ is Rashba coefficient [10] with the relationship of $k_{SO} = \dfrac{m^*\alpha}{\hbar^2}$ [20].

The role of delta function is to trigger spin & momentum locking [10]. We set the delta functions as $\delta(S_\pm) = \mp 1$ so that the spin-up (or spin-down) state of electron prefer to choose +k (or –k) direction, in order to minimize the total energy. While the Rashba field $B_{Rashba}$ behaves as Lorentz-like force $q(v_x \times B_{Rashba})$ to bend the trajectory of the negatively charged particle $q$ [14], the vertical velocity $v_{\pm y}$ of electron is required to consider an additional acceleration term $a_y = q(v_x \times B_{Rashba})/m^*$. When the scattering time of electron is imported, the vertical displacement of electron is integrated by $s_{\pm y} = \int v_{\pm y}\, dt$ numerically at equilibrium.

To initialize the status of Monte Carlo simulation, the spin and location of electrons are assigned in random manner as shown in Figure 1.

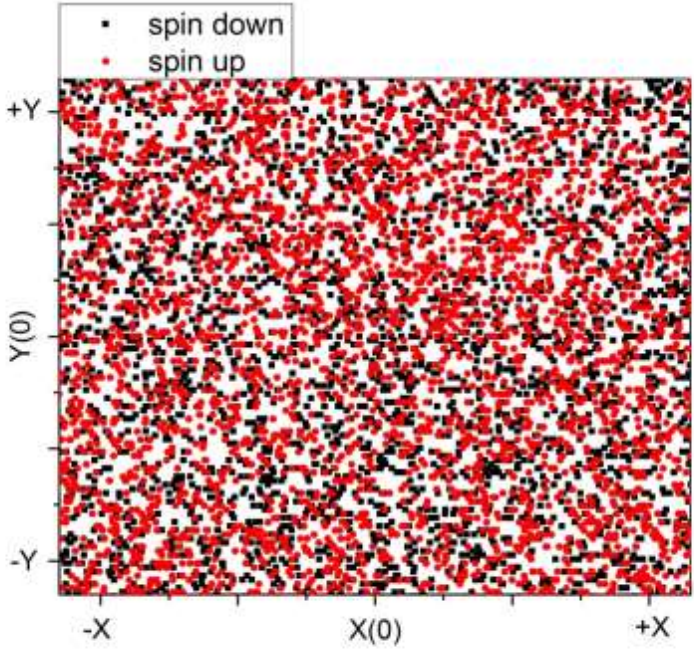

Figure 1: The spin state and location of electrons are randomized initially.

At each Monte Carlo step, the randomly selected electron may change the direction of spin and motion [13]. The selected electron is subjected to a trial state where the electron reverses the direction of spin if the random number ($0 \leq R_1 \leq 1$) is greater than 0.5. Otherwise, the spin state of electron retains. After the spin state is registered, the direction of motion along x direction is controlled by another random number $0 \leq R_2 \leq 1$ [15]. If $R_2$ is greater than 0.5, the electron moves towards +x direction [15]. Otherwise, the electron drifts to opposite direction. In the presence of Rashba spin-orbital coupling, the modified wavevector of electron is $k_T = k \pm k_{SO}$ [10].

When the modified wavevector is multiplied by $\frac{\hbar \tau}{m^*}$, the range of movement is attained approximately where $\tau$ is the average scattering time of electron.

The trial step is accepted if the relative Hamiltonian $\Delta H$ is less positive [13,15]. In addition, the spin-flip process and kinetic energy of electron are monitored by Boltzmann factor $e^{-\frac{\Delta H}{k_B T}}$ where $k_B$ is Boltzmann constant [13,15]. The trial state is accepted alternatively if the random number $0 \leq R_3 \leq 1$ is smaller than the Boltzmann factor [13,15]. The iterative process continues until the system reaches equilibrium state. The Rashba coefficient at finite temperature is modified by $\alpha' = \alpha + 1.75 \times 10^{-4} T$ [16,17]. We define the ideal Rashba effect as follow:

(1) No electron is found in the interior area.

(2) Electrons with different spin states occupy in two opposite edges. We name it as Rashba edges.

(3) The ratio $R_s$ between the spin-up state and spin-down state is within 0.98 < $R_s$ < 1.02.

In this article, Our Monte Carlo algorithm is applied to investigate the behavior of the Rashba effect of Graphene/Nickel (111) substrate [9]. Table 1 shows the simulation parameters.

Table 1: The simulation parameters of Graphene/Ni(111) substrate

| | |
|---|---|
| Fermi energy $E_F$ | ~4 eV [18] |
| Rashba wavevector $k_{SO}$ (Å) | 0.08 Å [9] |
| Exchange energy J | -0.014 eV [18] |
| Scattering time t | 20 fs [19] |
| Average effective mass $m^*$ | ~0.6 x $10^{-31}$ kg [18] |
| Rashba coefficient $\alpha_0$ | 1 eVÅ [9] |

Results:

Figure 2 shows the spatial distribution of electron in Graphene/Nickel(111) substrate at 0.01K where an ideal Rashba effect is observed. When temperature is increased to 20K, the non-ideal Rashba effect is observed in Figure 3 where some of the spin-polarized electrons are escaped from Rashba edges. Despite an electron with spin-down state is noticed in the upper half-plane, most of the electrons in the upper half-plane are occupied at spin-up state. In 64 x 64 square lattice, there are 4096 electrons in which 36 electrons are escaped from Rashba edges in Figure 3. While 0.8% electrons are escaped from Rashba edges at 20K, the probability of ideal Rashba effect in Graphene/Nickel (111) substrate equals to 99.2%.

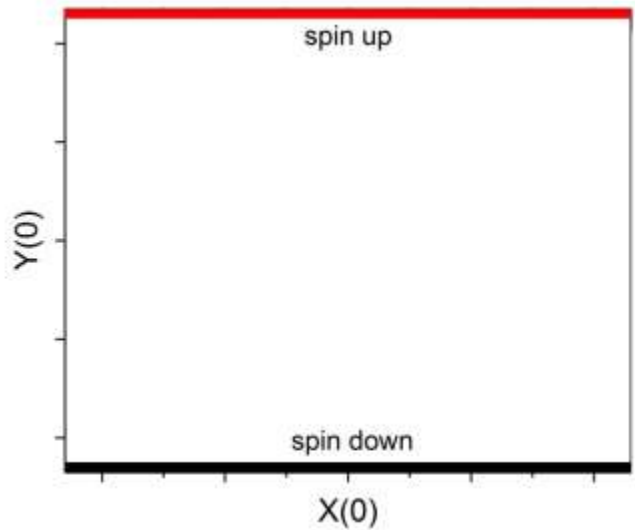

Figure 2: The spin-up state and spin-down state of electrons are accumulated in two opposite edges (Rashba edges).

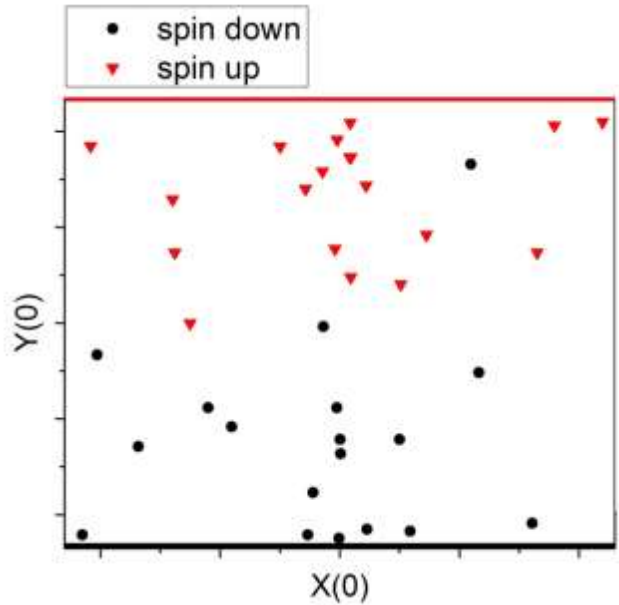

Figure 3: The distribution of spin and location of electrons in Graphene/Nickel (111) substrate at 20K.

To examine how thermal energy diminishes the Rashba spin-splitting phenomenon, we show the probability of ideal Rashba effect as a series of temperatures in Figure 4. At temperature as low as 0.1K, Graphene/Nickel (111) substrate (α = 1eVÅ) achieves an ideal Rashba effect with the probability of 0.999. Even the composite is heated up to 300K, the probability function still remains ~90% [9]. However, the probably function is very sensitive to Rashba coefficient. When the arbitrarily generated Rashba coefficient (α = 0.1eVÅ) is used, the probability of ideal Rashba effect is dropped dramatically upon heating where ~70% electrons are escaped from Rashba edges at 300K.

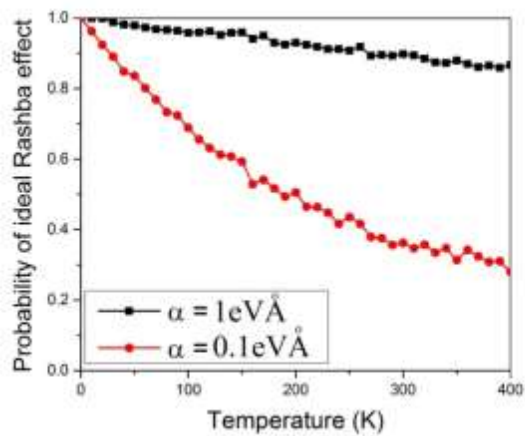

Figure 4: Probability of ideal Rashba effect in Graphene/Nickel (111) substrate as a function of temperatures with different Rashba coefficients

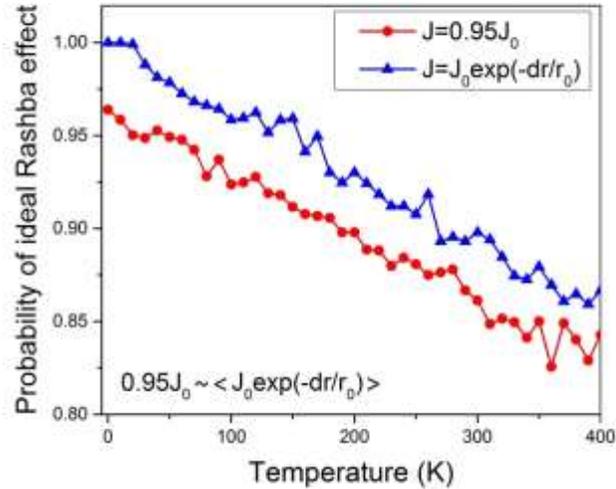

Figure 5: Probability of ideal Rashba effect in Graphene/Nickel (111) substrate as a series of temperatures. The red curve refers to a situation where the ferromagnetic proximity interaction is spread constantly inside Graphene. The blue curve refers to a more realistic case in which the electrons in Graphene may obtain non-uniform ferromagnetic proximity interaction depending on its spatial distribution.

We study the coupling factor between Rashba spin-orbital coupling and ferromagnetic proximity interaction (RF factor) in Figure 5. An interesting phenomenon is that the removal of the fluctuation term $e^{-\frac{dr}{r_0}}$ destabilizes the Rashba spin-splitting phenomenon. Despite the average strength of these two exchange couplings match each other, the electrons exposed to a non-uniform ferromagnetic proximity interaction reinforce the Rashba spin-splitting phenomenon. Figure 5 confirms that the rate of ferromagnetic fluctuation does not affect the slopes of the probability function. Apparently, the ferromagnetic fluctuation is beneficial to boost the layer-to-layer coupling [9,10]. Figure 6 demonstrates the Rashba spin-splitting phenomenon in Graphene / Nickel(111) substrate at 150K. About 97% electrons are stayed on Rashba edges. When thermal energy begins to randomize the spin and location of electrons at high temperatures, the distribution of electron shows Gaussian-like shape in Figure 6, where the Gaussian peak are located at X(0) regardless of spin state. The Gaussian-like peaks may be unobvious in Figure 3 but the peaks are emerged clearly in Figure 6

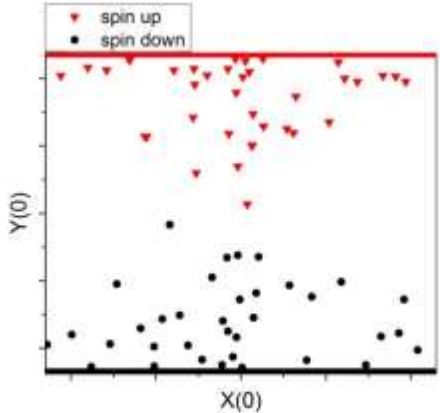

Figure 6: The distribution of spin and location of electrons in Graphene/Nickel (111) substrate at 150K.

Discussion:

Our Monte Carlo approach is able to produce an ideal spin-splitting phenomenon of Rashba effect at low temperatures with the evidences of Figure 2. While the magnitude of Rashba potential is symmetric on the upper and lower half plane, the electrons are divided into spin-up state and spin-down state in equal partition regardless of temperatures [3,8,9,10]. Though the spin-down state is more preferable to be appeared in the lower-half plane, Figure 3 displays that there is one electron with spin-down state occupying in the upper half-plane. This unexpected spin-down state in the upper-half plane is due to thermal fluctuations [13,15]. The thermal fluctuation in our code is determined by the competition between random number $R_3$ and Boltzmann factor [13,15]. If the sample is at 0K, the Boltzmann factor is always smaller than $R_3$ and therefore the randomization of spin [13] and location [15] of electron are rare. In contrast, the spin-polarized electron may accept the trial state and move to interior area if Rashba potential is masked by thermal energy [10].

Our Monte Carlo algorithm works reasonably to predict the room-temperature Rashba effect in Graphene / Ni(111) substrate [9] with the evidence of Figure 4. The simulation parameters in Table 1 allow us to simulate the experimental situation that nearly all electrons occupy on Rashba edges at room temperature [9]. The thermal softening of spin-orbital coupling [14] causes some of the electrons which move from Rashba edges to the interior area of Graphene / Nickel (111) substrate. However, the probability function never drops to zero because electrons are always found on the edges even Rashba effect does not exist. The thermal dependence of Rashba coefficient $\alpha' = \alpha + 1.75x10^{-4}T$ may not applicable above 400K [16,17] and therefore we do not simulate Rashba effect above 400K. Our Hamiltonian is suitable to study the Rashba effect of carbon-based materials because the atomic number of carbon (Z=6) does not exhibit strong intrinsic spin-orbital coupling [14] which allows us to describe Hamiltonian in three terms only.

Not all 2D materials emerge room-temperature Rashba effect under ferromagnetic substrates [1,9,10]. One of the reasons can be interpreted in Figure 5. The Rashba spin-splitting phenomenon is reinforced if the exchange energy is not a constant as a function of space and hence the strong Rashba spin-orbital coupling in Graphene / Nickel (111) composite is credited to the resultant rate of exchange fluctuation. Even nickel is replaced by another strong ferromagnetic substrate X, the Rashba spin orbital coupling may be weaker if the resultant rate of exchange fluctuation in Graphene / X substrate is very slow. In view of this, we have to take the resultant rate of exchange fluctuation into account in order to enhance the Rashba spin orbital coupling in carbon-based materials, instead of tuning the strength of ferromagnetism in substrate only. Based on this argument, introducing a tiny amount of vacancy defects in Graphene / Nickel (111) substrate may actuate the rate of exchange fluctuation and therefore the Rashba spin-splitting phenomenon of the defected Graphene / Nickel (111) substrate may be enhanced even stronger [21]. We draw a similar conclusion to the Density Functional Calculations computed by Shifei Qi *et al* [21] where the giant Rashba spin-orbit splitting in Graphene should be observed with help of co-doping [21].

To address why the escape route of electron forms Gaussian-like peak at X(0) at finite temperature in Figure 6. The Rashba field behaves as magnetic field to split the trajectory of spin-polarized electrons [10]. Despite our imported Rashba coefficient is spatial independent, the electrons bombard the walls along y-axis more vigorously that increases the local rate of exchange fluctuation. Based on Figure 5, the Rashba spin-splitting phenomenon is strengthened by exchange fluctuations. As a result, the Rashba spin-splitting phenomenon nearby the walls is more difficult to be destroyed by thermal energy that triggers the formation of Gaussian-like peaks at X(0). In view of this, spintronic engineers should avoid probing the spin state at X ~ 0 if Graphene / Ni (111) substrate is used as spintronic device at high temperatures. While the relationship between the Rashba k wavevector and exchange coupling of Graphene / Ni (111) substrate is still an open question, we do not tune the Rashba effect as a function of exchange coupling.